\documentclass[sigplan,screen,hyphens]{acmart}

\AtBeginDocument{%
  }

\setcopyright{none}
\renewcommand\footnotetextcopyrightpermission[1]{}
\acmConference{}{}{}

\acmISBN{}

\usepackage{multirow}

\begin{document}

\title{ESS: An Offload-Centric Latent-Cache Management Architecture for DeepSeek-V3.2-Exp}


\author{Xinhang Chen}
\email{chenxinhang@baidu.com}
\affiliation{%
  \institution{Baige AI Team, Baidu Inc.}
  \city{}
  \country{}
}

\author{Chao Zhang}
\email{zhangchao71@baidu.com}
\affiliation{%
  \institution{Baige AI Team, Baidu Inc.}
  \city{}
  \country{}
}

\author{Jiahuan He}
\email{hejiahuan@baidu.com}
\affiliation{%
  \institution{Baige AI Team, Baidu Inc.}
  \city{}
  \country{}
}

\author{Wei Liu}
\authornote{Wei Liu is the corresponding author.}
\email{liuwei88@baidu.com}
\affiliation{%
  \institution{Baige AI Team, Baidu Inc.}
  \city{}
  \country{}
}

\author{Jianming Zhang}
\email{zhangjianming@baidu.com}
\affiliation{%
  \institution{Baige AI Team, Baidu Inc.}
  \city{}
  \country{}
}

\author{Wenlong Zhou}
\email{zhouwenlong01@baidu.com}
\affiliation{%
  \institution{Baige AI Team, Baidu Inc.}
  \city{}
  \country{}
}

\author{Xiao Li}
\email{lixiao31@baidu.com}
\affiliation{%
  \institution{Baige AI Team, Baidu Inc.}
  \city{}
  \country{}
}

\author{Pai Zeng}
\email{zengpai@baidu.com}
\affiliation{%
  \institution{Baige AI Team, Baidu Inc.}
  \city{}
  \country{}
}

\author{Shiyong Li}
\email{lishiyong@baidu.com}
\affiliation{%
  \institution{Baige AI Team, Baidu Inc.}
  \city{}
  \country{}
}

\author{Yuanpan Qian}
\email{qianyuanpan@baidu.com}
\affiliation{%
  \institution{Baige AI Team, Baidu Inc.}
  \city{}
  \country{}
}

\author{Dong Li}
\email{lidong13@baidu.com}
\affiliation{%
  \institution{Baige AI Team, Baidu Inc.}
  \city{}
  \country{}
}

\author{Zhaogeng Li}
\email{lizhaogeng01@baidu.com}
\affiliation{%
  \institution{Baige AI Team, Baidu Inc.}
  \city{}
  \country{}
}

\renewcommand{\shortauthors}{}

\begin{abstract}
DeepSeek-V3.2-Exp introduces a sparse attention mechanism that significantly reduces inference latency in long-context scenarios. Although the overall throughput has improved greatly, the Decode-stage of PD disaggregation remains to be a major bottleneck. This bottleneck primarily stems from the conflict between linear growth of Latent-Cache with sequence length and the limited GPU memory capacity, which constrains the feasible batch-size and thereby suppresses Decode-stage throughput.

To address this challenge, we propose ESS (Extended Sparse Server), an offload-centric system design tailored for DeepSeek-V3.2-Exp. ESS selectively offloads Latent-Cache to CPU memory while preserving latency-critical components on GPU. By freeing up GPU memory, ESS effectively decoupling batch-size scaling from GPU memory constraints. This design significantly improves Decode-stage throughput, thereby reducing deployment costs in real-world settings.

Our high-fidelity simulations show that ESS delivers 69.4\% throughput improvement at 32K context length and up to 123\% throughput improvement at 128K, demonstrating its effectiveness for large-context inference workloads. These results highlight ESS as a practical and scalable solution for long-context LLM serving.
\end{abstract}



\maketitle

\section{Introduction}
The rapid growth of large language models (LLMs) \cite{achiam2023gpt, touvron2023llama, jiang2023mistral} has led to an increasing demand for serving long-context requests in various real-world applications, such as document summarization \cite{goyal2020evaluating, zhang2024benchmarking}, reasoning \cite{wei2022chain, liu2023evaluating}, code completion \cite{roziere2023code, chen2021evaluating}, dialogue systems \cite{thoppilan2022lamda, yuan2022wordcraft, wei2022emergent, zhang2024benchmarking} and agent-based decision making \cite{park2023generative, jiang2024casevo}. DeepSeek-V3.2-Exp \cite{liu2025deepseek}, in particular, adopts a sparse attention mechanism that substantially reduces the inference latency for these long-context scenarios. Despite these algorithmic optimizations, we observed that the Decode stage becomes a significant bottleneck under the PD disaggregation architecture \cite{pope2023efficiently}, in which Prefill and Decode are executed on disaggregated resources, especially as the sequence length increases.

The fundamental challenge lies in the memory footprint of the Latent-Cache. During Decode, the size of Latent-Cache increases linearly along with the processed sequence length. However, GPU memory capacity remains fixed, preventing the system from scaling batch-size as context length grows. This mismatch suppresses Decode-stage throughput and leads to rapidly increasing serving cost for long-context applications. Although sparse attention effectively alleviates LLM's computational burden, the GPU memory bottleneck imposed by Latent-Cache persists as the primary limitation for higher throughput.

Motivated by these observations, we aim to design a serving architecture specifically for DeepSeek-V3.2-Exp that significantly improves decode throughput without compromising the model's intrinsic accuracy.

Our key observation is that the Top-2K Latent-Cache selected by DeepSeek-V3.2-Exp exhibits strong temporal locality. This allows us to maintain a Sparse Memory Pool on the GPU that stores a subset of Latent-Cache entries and updates it dynamically during the Decode-stage. Based on this finding, we conducted a systematic bottleneck analysis to further verify our assumption. Eventually, we proposed ESS, a system design that extends effective GPU memory capacity through efficient Offload-Prefetch strategies, while preserving the sparse attention computation pipeline.

\begin{table}
  \caption{Basic Setting}
  \label{tab:setting}
  \begin{tabular}{cl}
    \toprule
    Optimization Options & Value \\
    \midrule
    MTP & 2 \\
    Node & 4 \\
    TP-Size & 1 \\
    EP-Size & 32 \\
    Context & 32K \\
    Attention-Engine & FlashMLA \\
    Two-Batch Overlap & open \\
    MTP-Accept-Ratio & 1.7 \\
    Model & DeepSeek-V3.2-Exp \\
    PCIe & 5th \\
    \bottomrule
  \end{tabular}
\end{table}

\sloppy
ESS addresses three fundamental challenges of Offload–Prefetch strategies. First, it leverages Unified Virtual Addressing (UVA) to mitigate the low bandwidth efficiency caused by small-granularity data transfers. Second, it reduces the total number of Cache Misses during the Decode phase through an LRU-based replacement policy and a corresponding warm-up mechanism. Third, it introduces a layer-wise overlap scheme that enables effective compute–communication pipeline. By resolving these bottlenecks, ESS significantly increases the achievable batch-size of long-context inference, thereby improving end-to-end throughput.

In this paper, our main contributions are as follows:
\begin{itemize}
\item \textbf{Feasibility analysis of Latent-Cache Offload for DeepSeek-V3.2-Exp}: We systematically evaluate the feasibility of Latent-Cache offloading in DeepSeek-V3.2-Exp, providing a comprehensive analysis of its performance, key bottlenecks, and achievable gains across various system configurations and context lengths.

\item \textbf{The ESS system}: ESS introduces an offload-centric architecture that enables lossless scaling of Decode-stage throughput while preserving model accuracy. By transferring Latent-Cache storage to the CPU side, ESS alleviates the memory-capacity constraints that typically limit Decode performance. It also works seamlessly with existing inference optimizations, such as MTP and Two-Batch overlap, making it a practical enhancement for large-scale industrial LLM serving systems.

\begin{figure}[h]
  \centering
  \includegraphics[width=\linewidth]{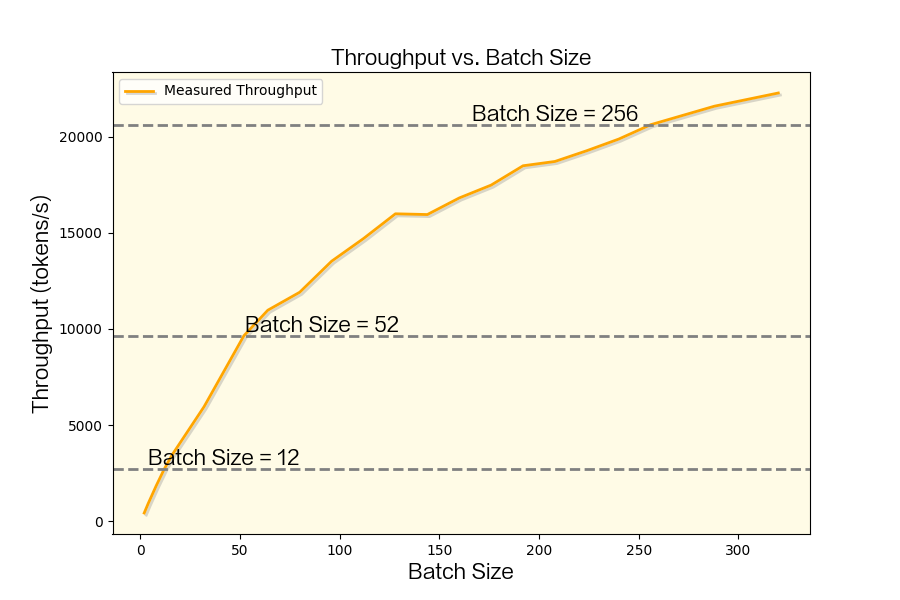}
  \caption{Relationship Between Throughput and Batch Size.}
  \label{fig:throughputVsBatchsize}
\end{figure}

\item \textbf{A high-fidelity simulator for reliable performance prediction}: We build a simulation framework that precisely models computation, communication, and Offload–Prefetch overheads while reconstructing the full execution pipeline based on real computation and data-transfer flows. The simulator also incorporates system-level optimizations such as MTP and Two-Batch overlap execution to capture their impact faithfully. With this high-precision modeling, we can accurately estimate the LLM inference performance without relying on extensive real-world experiments, thereby substantially reducing validation costs and accelerating system design iteration.
\end{itemize}

Simulations show that ESS improves throughput by 69.4\% at 32K and by up to 123\% at 128K, demonstrating robust scalability under long-context workloads..

\begin{figure*}[h]
  \centering
  \includegraphics[width=\linewidth]{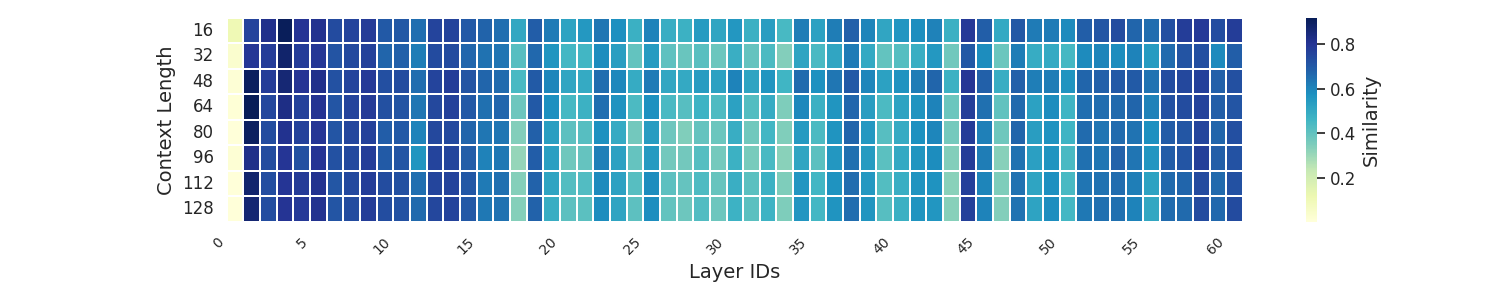}
  \caption{Intra-Layer Similarity Across Different Context Lengths.}
  \label{fig:intra-layer-similarity}
\end{figure*}

\section{Background and Motivation}

\subsection{GPU Memory Limits Throughput}
Figure \ref{fig:throughputVsBatchsize} illustrates the relationship between batch-size and Decode throughput under a 32K context length, with all data generated by our high-fidelity simulator; the corresponding system configuration is shown in Table \ref{tab:setting}. In theory, throughput should continue to increase as batch-size grows. This is primarily because batch-size is closely tied to the arithmetic intensity of GEMM operations. Larger batches yield higher Model FLOPs Utilization (MFU) and thus improve overall compute utilization.

However, with our current configuration, we observe that GPU memory capacity becomes the critical bottleneck: the batch-size can only scale up to 52, achieving a maximum throughput of just 9,647 tokens/s. Due to the full occupation of GPU memory, the system cannot further increase the computation batch, resulting in a throughput that is far below the hardware theoretical upper-bound. From this, we conclude that GPU memory capacity is the primary factor blocking Decode-throughput scaling.

This finding highlights the necessity and value of Offload-Prefetch strategies: by decoupling the Latent-Cache from GPU memory, the system can break through the existing throughput ceiling and unlock its computational potential.

\subsection{Temporal Locality of Important Cache Entries}

To alleviate GPU memory pressure while preserving model accuracy, offloading part of the Cache to the CPU is a practical and feasible optimization strategy. However, for this approach to be effective, a critical prerequisite must be satisfied: the Cache access pattern must exhibit strong locality. Only when the model’s access pattern to the Latent-Cache exhibits sufficient repetitiveness or locality can a high cache hit rate be sustained. Otherwise, extensive CPU–GPU data transfers over PCIe will introduce substantial latency, ultimately negating the benefits of the Offload-Prefetch strategy.

To determine whether the Latent-Cache in DeepSeek-V3.2-Exp exhibits sufficient locality, we analyze its access pattern from the perspective of intra-layer access. This metric captures the stability and consistency of Cache indices accessed within the same layer between consecutive decoding steps.

We define a metric, \textbf{Intra-Layer Similarity}, to quantify the locality of the two access patterns described above. Let $I^l$ denote the set of Latent-Cache indices currently stored on the GPU for layer $l$, and let $K_t^l$ represent the Top-K index set required by layer $l$ during the decoding step $t$. Equations (\ref{eq:intra-layer-similarity}) provide the mathematical definitions of intra-layer similarity. This metric is constructed based on set similarity, where a larger overlap indicates better locality and thus provides more favorable conditions for the effectiveness of the Offload-Prefetch strategy.

\begin{equation}
\label{eq:intra-layer-similarity}
  r_t^l=\frac{|K^{l}_{t-1} \cap K^l_t |}{|K^l_t|}
\end{equation}

Based on the pattern evaluation of LongBench V2 to DeepSeek-V3.2-Exp (see Figure \ref{fig:intra-layer-similarity}), we observe consistent and high similarity in intra-layer access patterns. This indicates that Latent-Cache accesses exhibit strong locality. Consequently, using CPU memory as an extension of HBM through offloading remains a viable approach. 

However, when integrating the PD disaggregation architecture and the SGLang inference framework with the DeepSeek-V3.2 model, we encounter several unique challenges that set our scenario apart from prior CPU–GPU collaborative storage efforts.

\begin{figure*}[h]
  \centering
  \label{fig:pd}
  \includegraphics[width=\linewidth]{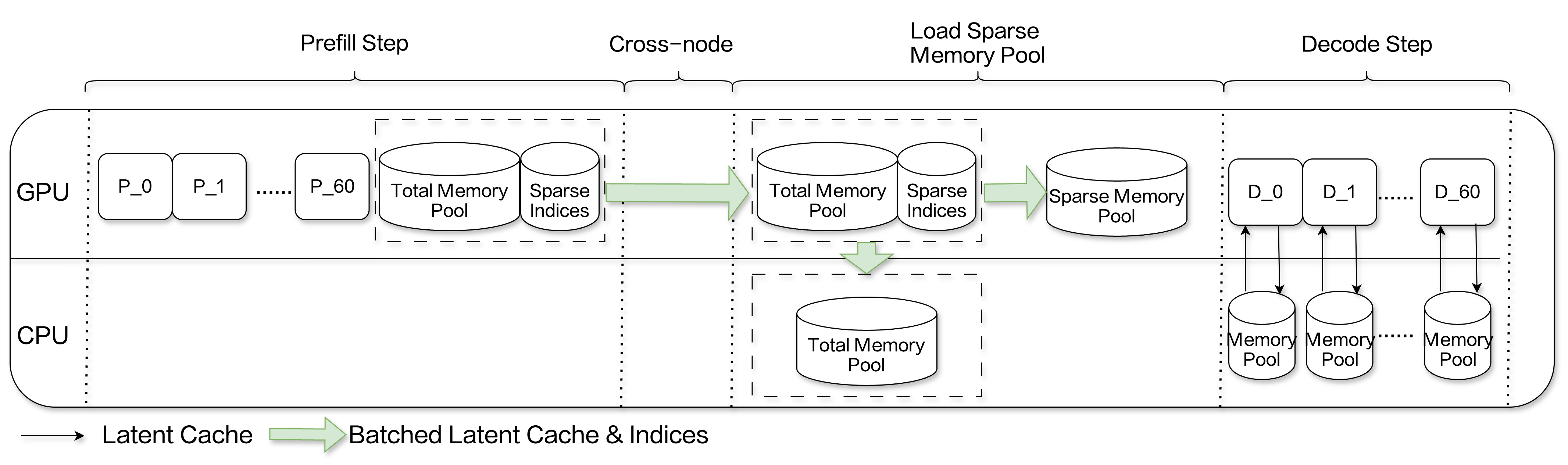}
  \caption{Latent-Cache Offload–Prefetch Timing in the PD disaggregation Setting.}
  \label{fig:PD}
\end{figure*}

\begin{itemize}
\item \textbf{Inefficient Small-Block Bandwidth}: In DeepSeek-V3.2-Exp, each Cache block is only 656 bytes, and the 2,048 Cache blocks accessed at each step are scattered across the Memory Pool. Such highly fragmented, small-block access patterns make it difficult for PCIe to form efficient batch transfers, substantially reducing bandwidth utilization and becoming a primary bottleneck for offload-based designs.

\item \textbf{Difficulty of Maintaining a Low Miss Count}: To increase the batch-size, the amount of Cache residing on the GPU must be minimized. However, reducing the GPU-side Cache raises the likelihood of Cache Misses, thereby increasing the volume of H2D (Host-to-Device) transfers. Since these transfers cannot be fully overlapped with computation, the resulting data-transfer latency causes the Decode-stage throughput to fall short of expectations.

\item \textbf{Unhidden Data-Transfer Latency}: During the Decode stage, the amount of computation available is insufficient to fully hide data-transfer latency. When Cache miss occurs frequently, these transfer delays become exposed along the critical path, further degrading the inference performance.

\end{itemize}

\section{ESS System Design and Analysis}
As discussed earlier, Offload–Prefetch is a widely adopted lossless optimization strategy in recent years, particularly suitable for inference scenarios that are highly sensitive to accuracy. Following the same principle, our design also places offloading at its core. For DeepSeek-V3.2-Exp, the Cache consists of two components: the Indexer-Cache and the Latent-Cache. The Indexer Cache is responsible for computing the importance of each Latent-Cache entry and selecting the top 2,048 critical entries for computation. According to our architectural analysis, the Indexer Cache requires full computation and accounts for only 16.8\% of the total Cache storage. Based on this observation, we choose not to offload the Indexer-Cache. Instead, we apply the Offload–Prefetch optimization solely to the Latent-Cache.

Figure \ref{fig:PD} summarizes the triggering points of offload and prefetch operations for the PD disaggregation architecture. In addition, this subsection further analyzes the key challenges outlined in Section 2.2 and introduces corresponding optimizations derived from these insights.

\subsection{Small-Block Data Transfers}
Although PCIe 5.0 provides up to 64 GB/s of unidirectional bandwidth, which makes Offload–Prefetch–style approaches theoretically feasible, this bandwidth is often difficult to fully utilize in modern inference frameworks such as SGLang. This is because SGLang employs PagedAttention \cite{kwon2023efficient}  to manage the Latent, partitioning it into multiple pages that are physically non-contiguous. Furthermore, DeepSeek-V3.2-Exp adopts an even finer-grained sparsification strategy, reducing the minimum swap unit to a single Latent-Cache entry. Such highly fragmented, small-block access patterns breaks the high-bandwidth PCIe link into numerous tiny transactions, drastically degrading the effective usable bandwidth.
Considering that each Latent-Cache block is only 656 bytes, the effective bandwidth of cudaMemcpyAsync in our measurements is extremely limited: \textbf{0.79 GB/s} for H2D and \textbf{0.23 GB/s} for D2H, respectively.

\subsubsection*{FlashTrans Accelerates Fragmented Data Transfers}

To mitigate this issue, we incorporate UVA into our design, allowing the GPU to directly access pinned memory on CPU side and reducing the management overhead associated with small-block transfers over PCIe. Utilizing UVA design, we design the FlashTrans operator, which enables address-based on-demand transfers and eliminates the scheduling overhead of frequent \texttt{cudaMemcpyAsync} calls. FlashTrans significantly improves effective bandwidth under fine-grained, non-contiguous Latent-Cache access patterns, enabling practical Offload–Prefetch in DeepSeek-V3.2-Exp. 
Our experiments show that FlashTrans achieves \textbf{37 GB/s} for H2D transfers and \textbf{43 GB/s} for D2H transfers.

\subsection{Cache Miss Behavior}
A low Cache Miss Count significantly reduces data-transfer volume. To achieve this, ESS employs an LRU-based swap engine that systematically analyzes Cache Miss behavior throughout the inference process. In addition, we introduce LRU-Warmup, a technique designed to maintain a low Miss Count during the initial stages of decoding.

\subsubsection*{LRU-Warmup}
\begin{figure*}[h]
  \centering
  \begin{minipage}{0.48\linewidth}
    \centering
    \includegraphics[width=\linewidth]{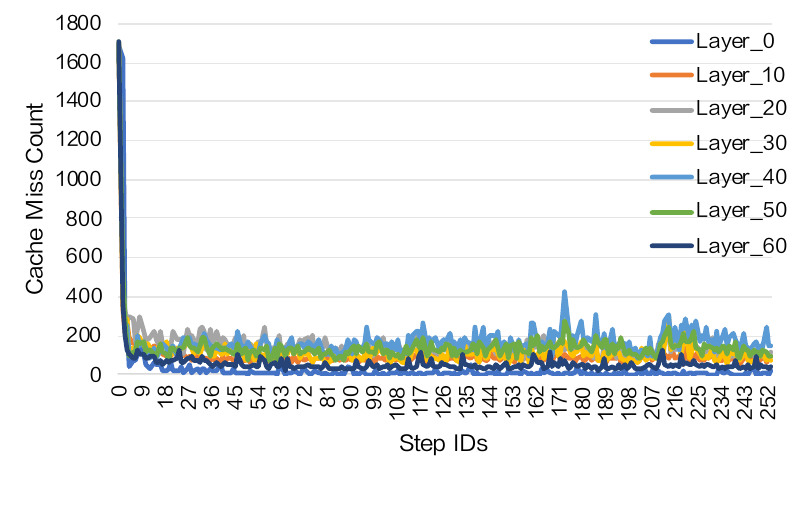}
    \caption*{Before LRU-Warmup}
  \end{minipage}
  \hfill
  \begin{minipage}{0.48\linewidth}
    \centering
    \includegraphics[width=\linewidth]{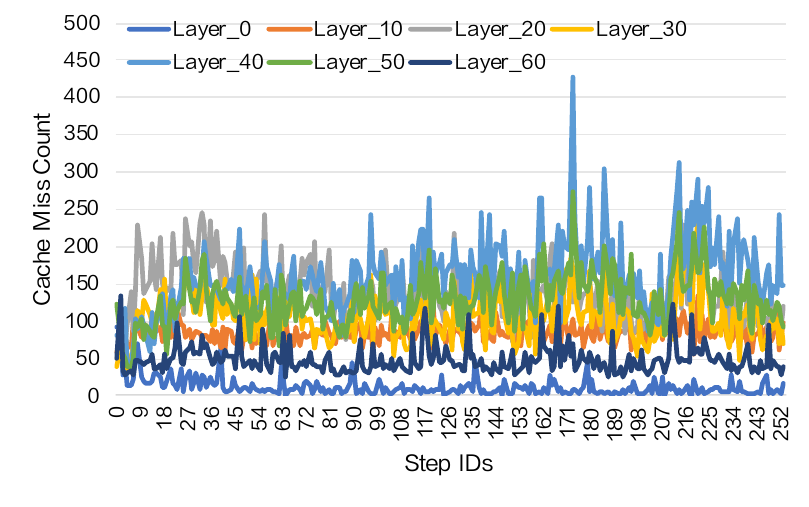}
    \caption*{After LRU-Warmup}
  \end{minipage}
  \caption{Effect of LRU-Warmup on Early Decode Cache Miss Count (MTP = 1, Sparse Memory Ratio = 0.2).}
  \label{fig:warmup_comparison}
\end{figure*}

The initial state of the GPU-side Sparse Memory Pool has a significant impact on the performance of early-stage generation. As shown in Figure \ref{fig:warmup_comparison}, a large number of Cache Miss occurs at the beginning of decoding, but this effect quickly diminishes as decode progresses. To reduce the additional overhead during this initial phase, we apply an LRU Warm-up to preheat the LRU Cache. Specifically, we extract the Top-2K Latent-Cache IDs selected within the last 32 windows of the Prefill stage and sequentially insert them into the LRU cache, thereby constructing a cache state that better matches the access requirements of early decoding. As shown in Figure \ref{fig:warmup_comparison}, this strategy substantially reduces the Cache Miss Count at the beginning of Decode, improving efficiency during the early stages of inference.

\subsubsection*{LRU-Based Cache Eviction and Admission}
In Section 2.2, we demonstrate that Latent-Cache access in DeepSeek-V3.2-Exp exhibits strong temporal locality both within layers and across layers. This property indicates that once a Latent-Cache entry is accessed at a given step, it is highly likely to be accessed again in subsequent steps. Motivated by this observation, we apply an LRU-Based strategy to continuously update the GPU-side Sparse Memory Pool, ensuring that entries with the highest likelihood of future reuse are preferentially retained.

\begin{figure}[h]
  \centering
  \includegraphics[width=\linewidth]{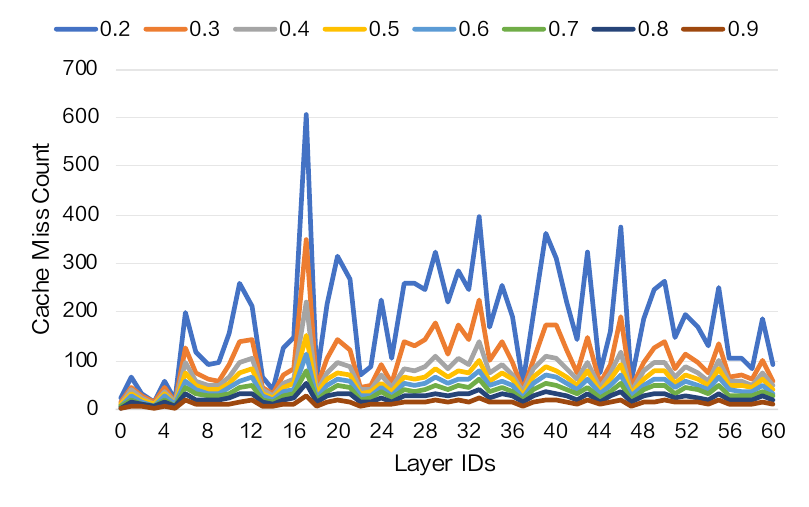}
  \caption{Intra-Layer Cache Miss Analysis.}
  \label{fig:cachemiss}
\end{figure}

Figure \ref{fig:cachemiss} reports the average number of Latent-Cache misses per batch when the Sparse Memory Pool is constructed based on Intra-Layer Access. We further compare different Sparse Memory Ratios, the proportion of the total Cache that resides on the GPU, and these results collectively characterize the Cache Miss Count under varying GPU memory budgets, helping us identify appropriate Sparse Memory Ratio values and reveal underlying patterns.

\subsection{Compute–Communication Overlap}
Another key factor that affects end-to-end performance is the overlap between computation and communication. Based on a systematic breakdown and analysis of the existing implementation in SGLang, we design two techniques, Dual-Attention (DA) Overlap and DualBatch-Attention (DBA) Overlap, to maximize the overlap between computation and data transfer, thereby further improving overall throughput.

\subsubsection*{Analysis of the Impact of Overlap}
\begin{figure*}[h]
  \centering
  \begin{minipage}{0.60\linewidth}
      \centering
      \includegraphics[width=\linewidth]{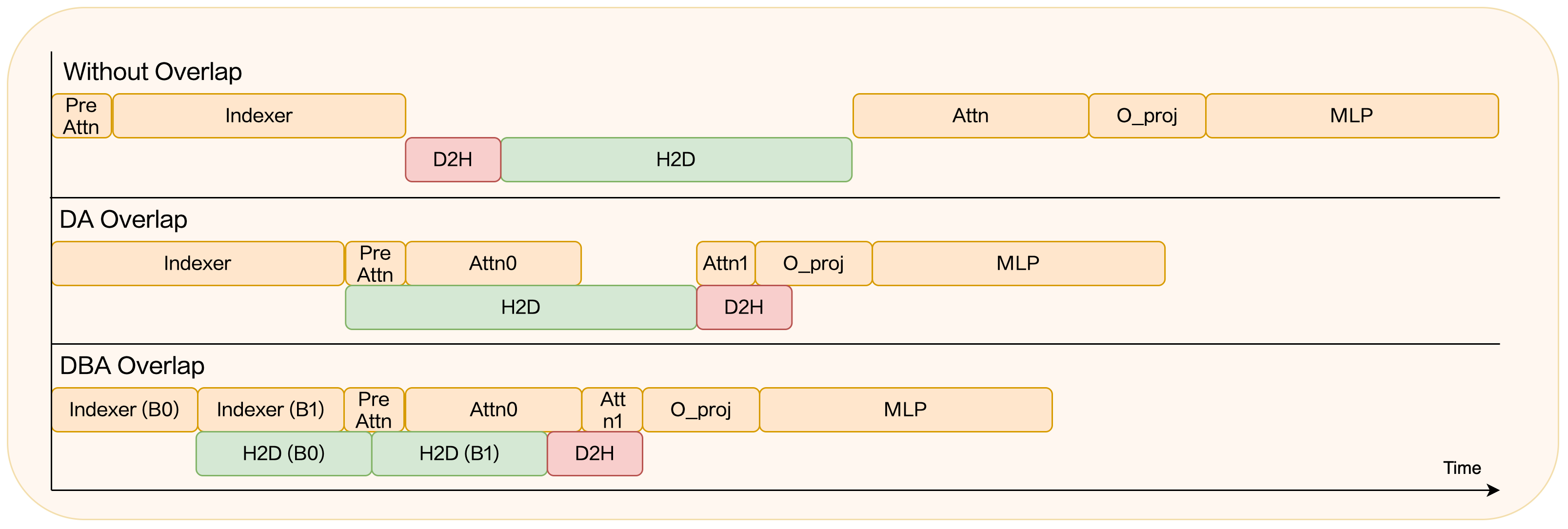}
      \caption{Comparison of Overlap Strategies.}
      \label{fig:overlap}
  \end{minipage}
  \begin{minipage}{0.36\linewidth}
    \centering
    \includegraphics[width=\linewidth]{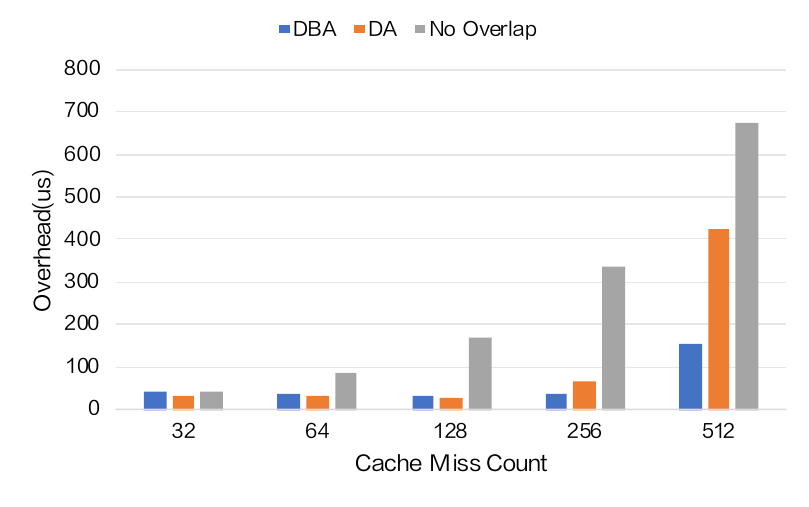}
    \caption{Overhead Comparison of Overlap Strategies.}
    \label{fig:overlap_comparision}
  \end{minipage}
\end{figure*}

Figure \ref{fig:overlap} shows the complete inference timeline when no Overlap strategy is applied. In the figure, H2D denotes the subsequent retrieval of the Latent-Cache. In addition, a smaller D2H operation is present, which writes the newly generated Latent-Cache from the current step back to the CPU-side Total Memory Pool.

Without Overlap enabled, both types of data transfer must wait for the Indexer computation to finish before they can begin, and the subsequent Attention computation cannot proceed until all transfers have completed. This strict dependency prevents any parallelism between stages, resulting in a fully serialized execution pipeline that significantly limits overall throughput.

\subsubsection*{Without Overlap} This mode corresponds to the default implementation in SGLang, where there's no overlap between computation and communication. In this setting, the GPU remains idle while waiting for H2D transfers of Latent-Cache data, resulting in markedly lower end-to-end throughput compared with the achievable hardware limits.

\subsubsection*{DA Overlap} \noindent In DeepSeek-V3.2-Exp’s SGLang implementation, Attention consists of two stages: \texttt{forward\_prepare} and \texttt{forward\_core}. The \texttt{forward\_prepare} stage can be further decomposed into PreAttn and Indexer. Indexer session represents both the Indexer itself and all computations that depend on its results, while PreAttn includes operations independent of the Indexer, such as \texttt{q\_b\_proj}, \texttt{bmm}, \texttt{copy\_pe}, \texttt{and rotary\_embedding}.

To improve the degree of overlap between computation and H2D prefetching, we first decouple PreAttn from \texttt{forward\_prepare} and defer its execution until after the Indexer finishes. However, the compute workload of PreAttn alone is insufficient to fully hide the Prefetch latency of Latent-Cache.
To further improve the overlap ratio, we divided SparseMLA into two sub-stages: Attn0 and Attn1. Attn0 proceeds Latent-Cache already resident on the GPU, while Attn1 waits for the complete of H2D prefetch and then continues using the newly fetched cache. The results of the two stages are then merged.
Since Attn0 can run concurrently with H2D transfers, this design effectively hides data transfer latency and significantly improves overall overlap efficiency.

\begin{figure}[h]
  \centering
  \includegraphics[width=\linewidth]{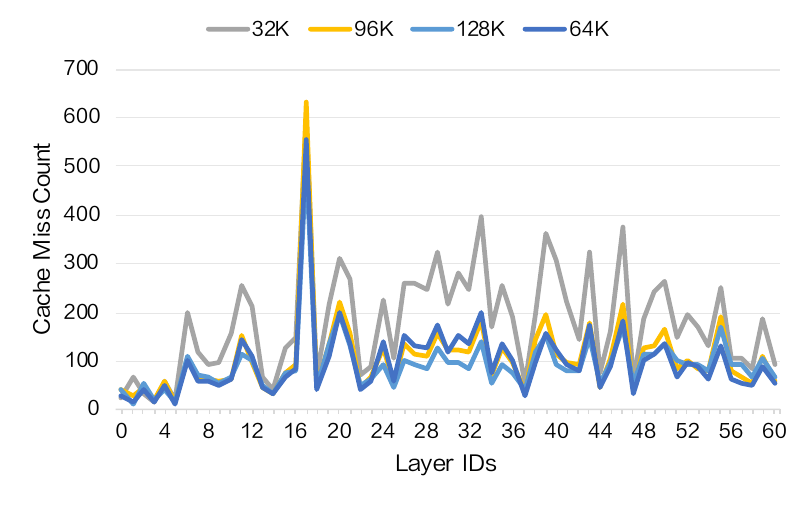}
  \caption{Cache Miss Count Comparison Across Different Context Lengths (MTP = 2, Ratio = 0.2).}
  \label{fig:cachemiss_parttern}
\end{figure}

\subsubsection*{DBA Overlap} Because the compute cost of Attention remains nearly constant once the context length exceeds 2K, the overlap benefit of Dual-Attn becomes limited in long-context scenarios. To further expand the opportunity of overlap, we propose DualBatch-Attention (DBA) Overlap.
DBA builds upon Dual-Attn by partitioning the Indexer along the batch dimension, enabling approximately half of the Indexer computation to participate in the overlap. This not only increases the amount of computation that can be hide but also preserves substantial compute–communication overlap even with long-context settings. As the result, we could further improve end-to-end throughput.

More specifically, in DBA Overlap, we primarily incorporate the \texttt{paged\_mqa\_logits} computation and the Top-K selection into the overlap region. We choose these components because their computational intensity does not decrease significantly when the batch-size is reduced. This allows them to effectively compensate for the performance loss introduced by batch partitioning and thereby improve the overall overlap efficiency.
Figure \ref{fig:overlap} presents a timeline comparison between DA Overlap and DBA Overlap during actual execution.

\subsubsection*{Layer-Wise Overlap Strategy}
As shown in Figure \ref{fig:cachemiss}, the Cache Miss behavior varies significantly between different layers, especially when the Sparse Memory Ratio is small. For example, when the Sparse Memory Ratio is 0.2, the number of Cache Misses per batch ranges from 16.66 to 605. This large variability indicates that a single Overlap strategy cannot effectively handle all layers.

Under the setting of Context = 128K, BS = 160, MTP = 2, Two-Batch Overlap enabled, and PCIe bandwidth = 37 GB/s, we evaluate the performance degradation of the three Overlap strategies under varying Cache Miss Counts, as illustrated in Figure \ref{fig:overlap_comparision}. In the DBA Overlap strategy, the computational cost of the Indexer increases linearly with the context length, enabling it to effectively hide Latent-Cache transfer latency in long-context scenarios. Consequently, DBA remains highly effective when the Cache Miss count reaches 512.
When the Cache Miss level is lower, DA Overlap becomes more favorable, as it can fully hide the data-transfer time without introducing the Indexer splitting overhead.

We argue that the choice of Overlap strategy is primarily determined by two factors: the Cache Miss pattern and the context length. First, we observe, that across different context lengths, the Cache Miss distribution along the Layer ID dimension exhibits highly consistent trends (as shown in Figure \ref{fig:cachemiss_parttern}). This allows us to identify, through offline profiling, the key layers that are most likely to incur a large number of Cache Misses during inference.
On the other hand, under the same Sparse Memory Ratio, the overall Cache Miss level varies with the context length. Therefore, it is necessary to empirically determine the Cache Miss threshold at which the system should switch to the DBA strategy.

\begin{figure}[h]
  \centering
  \includegraphics[width=\linewidth]{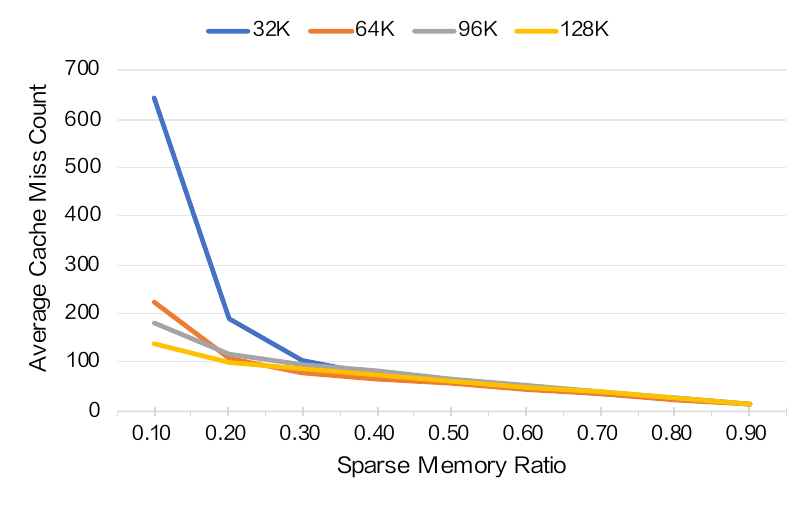}
  \caption{Cache Miss Behavior Across Different Context Lengths.}
  \label{fig:varyingContextLen}
\end{figure}

\subsection{Scalability Across Various Context Lengths}
As shown in Figure \ref{fig:varyingContextLen}, along with the increase of context length, the average Cache Miss Count remains relatively stable with Sparse Memory Ratio greater than 0.2. Noticeably, the most severe Cache Miss behavior occurs in the 32K context setting when the Sparse Memory Ratio is small. This is primarily due to the limited Sparse Memory Pool size, which triggers frequent swap-in and swap-out operations and substantially increases the likelihood of misses. From a scalability perspective, we recommend configuring the GPU buffer to be no smaller than 6.4K.

Furthermore, Figure \ref{fig:varyingContextLen} also indicates that, under the same Sparse Memory Ratio, the average Cache Miss Count decreases as the context length grows. This suggests that ESS yields even higher throughput improvements in longer-context scenarios.

\section{Simulation Validation}
\subsection{Simulator}
We conduct our performance evaluation using an internally developed high-fidelity simulator. The simulator is based on metadata collected from real machine executions, with missing data points filled in through linear interpolation. It constructs the full execution pipeline based on actual computation and data-transfer flows, while also incorporating system-level optimizations such as MTP and Two-Batch Overlap execution.
Thanks to its high-accuracy modeling, the simulator provides a reliably prediction of the inference performance of large models without requiring extensive real-world experiments, that significantly reduced the cost of design validation.

\begin{table}[t]
\centering
\small
\renewcommand{\arraystretch}{1.25} 

\begin{tabular}{c|c|c|c}
\hline
\textbf{Setting} & \textbf{Batch Size} & \textbf{Throughput} & \textbf{OTPS} \\
\hline

\multirow{5}{*}{
\begin{tabular}{@{}c@{}}
\textbf{MTP = 2} \\
\textbf{Context = 32K} \\
\textbf{Accept Ratio = 1.7}
\end{tabular}}
 & 52 (1)  & 9647.71  & 23.19 \\
 & 64 (0.82) & 10693.31 & 20.89 \\
 & 96 (0.48) & 13155.98 & 17.13 \\
 & 128 (0.31) & 15620.14 & 15.25 \\
 & 160 (0.21) & 16347.88 & 12.77 \\
\hline

\multirow{5}{*}{
\begin{tabular}{@{}c@{}}
\textbf{MTP = 4} \\
\textbf{Context = 32K} \\
\textbf{Accept Ratio = 2.8}
\end{tabular}}
 & 52 (1) & 12168.02 & 29.25 \\
 & 64 (0.82) & 13656.66 & 26.67 \\
 & 96 (0.48) & 15814.07 & 20.59 \\
 & 128 (0.31) & 17746.10 & 17.33 \\
 & 160 (0.21) & 17601.03 & 13.75 \\
\hline

\multirow{5}{*}{
\begin{tabular}{@{}c@{}}
\textbf{MTP = 4} \\
\textbf{Context = 32K} \\
\textbf{Accept Ratio = 3.4}
\end{tabular}}
 & 52 (1) & 14775.45 & 35.52 \\
 & 64 (0.82) & 16583.08 & 32.39 \\
 & 96 (0.48) & 19202.80 & 25.00 \\
 & 128 (0.31) & 21548.83 & 21.04 \\
 & 160 (0.21) & 21372.68 & 16.70 \\
\hline

\multirow{3}{*}{
\begin{tabular}{@{}c@{}}
\textbf{MTP = 2} \\
\textbf{Context = 128K} \\
\textbf{Accept Ratio = 1.7}
\end{tabular}}
 & 13 (1) & 3669.19 & 23.19 \\
 & 40 (0.2) & 6925.06 & 21.64 \\
 & 54 (0.1) & 8169.60 & 18.91 \\
\hline
\end{tabular}

\caption{Throughput and OTPS Under Different MTP and Acceptance Settings (Values in parentheses indicate the corresponding Sparse Memory Ratio).}

\label{tab:end2end_benefit_32k}
\end{table}

\subsection{End-to-End Performance Evaluation}
In this experiment, we evaluate the performance of the 32K context setting with different MTP values and acceptance rates, while keeping all other configurations consistent with Table \ref{tab:setting}. Based on the throughput and OPTS results produced by our simulator (Table \ref{tab:end2end_benefit_32k}), we observe that MTP = 2 achieves a 69.4\% improvement in end-to-end throughput. When MTP = 4 with an acceptance rate of 3.4, the end-to-end throughput improves by 45.8\%.

We further evaluate the 128K ultra-long context setting with MTP = 2 and an acceptance rate of 1.7. Since the batch-size remains relatively small at this context length, we disable the Two-Batch Overlap optimization in this experiment. As shown in Figure \ref{fig:varyingContextLen}, longer context lengths allow the system to operate with a lower Sparse Memory Ratio. The final results are summarized in Table \ref{tab:end2end_benefit_32k}, where a Sparse Memory Ratio of 0.1 yields a 123\% improvement in end-to-end throughput.

\section{Ralated Work}

In this section, we categorize prior work on KV-cache optimization into two main directions.

\paragraph{Static KV-Cache Compression} 
This line of work focuses on permanently removing less important KV entries based on predefined or pre-trained importance metrics.
H2O \cite{zhang2023h2o}, for example, computes a historical KV-score that reflects each token’s long-term contribution and retains only the most influential cache elements. SnapKV \cite{li2024snapkv} further exploits locality and reduces the cost of computing importance scores by maintaining an observation window that quickly identifies high-value KV entries. StreamingLLM \cite{xiao2023efficient} observes a different phenomenon—KV activations tend to concentrate on the earliest few tokens, an effect referred to as the sink phenomenon, enabling aggressive pruning of later positions.

Other studies highlight that sparsity patterns vary significantly across layers and even across attention heads. PyramidKV \cite{cai2024pyramidkv} leverages this observation by allocating cache budgets in a layer-wise manner. AdaKV \cite{feng2024ada} goes one step further by adopting an adaptive strategy that assigns adaptive cache budgets to different heads based on their estimated utility. DuoAttention \cite{xiao2024duoattention} splits attention heads into streaming heads and retrieval heads: the streaming heads use a sink-plus-local attention pattern, while the retrieval heads compute full attention. This design substantially reduces overall attention complexity while preserving long-range modeling capability.

\paragraph{Dynamic KV-Cache Compression}
Unlike static compression methods, dynamic approaches generally aim to retain the entire KV cache, but during decoding they compute attention only over a selected subset of KV entries. For example, Quest \cite{tang2024quest} treats adjacent KV-Cache as clusters and, at each decoding step, selects different clusters to participate in attention computation, thereby reducing the overall computational cost. MagicPig \cite{chen2024magicpig} instead uses locality-sensitive hashing to efficiently retrieve important KV entries while offloading most of the cache to CPU memory, which improves batch-size and throughput. RetrievalAttention \cite{liu2024retrievalattention} exploits the dynamic sparsity inherent in the attention mechanism, constructing Approximate Nearest Neighbor Search (ANNS) indexes for KV vectors in CPU memory and retrieving the most relevant KV pairs via vector search during generation. Other works following similar dynamic selection ideas include ShadowKV \cite{sun2024shadowkv}, PQCache \cite{zhang2025pqcache}, and FreeKV \cite{liu2025freekv}.

\paragraph{Offload-Prefetch}
This is an engineering-oriented optimization technique that is often combined with dynamic compression. The approach offloads a large portion of the KV-Cache to CPU memory and retrieves the required segments over PCIe only when needed for computation. FlexGen\cite{sheng2023flexgen} delivers high-throughput inference for billion-scale LLMs on a single commodity GPU by optimizing the placement of weights, activations, and KV-Cache across GPU–CPU–SSD, together with a block-level execution schedule and overlapped I/O. SparseServe \cite{zhou2025sparseserve} further addresses fragmented data transfers and employs dynamic batch-size management to prevent cache thrashing and excessive KV-Cache loading.

\section{Conclusion and Future Work}
As an engineering-oriented technique for increasing batch-size without compromising accuracy, Offload–Prefetch has been validated and widely adopted in many previous works. In this paper, we design and simulate the corresponding Offload strategy tailored to the inference pipeline of DeepSeek-V3.2-Exp using SGLang. Our experimental results demonstrate the feasibility and performance potential of Offload–Prefetch for this model, laying the groundwork for further system optimizations. In future work, we plan to integrate this approach into the actual framework and explore combining it with lossy compression methods such as SnapKV to further improve overall inference throughput.

\bibliographystyle{ACM-Reference-Format}
\bibliography{custom}

\appendix
\end{document}